\documentclass[aps,prd,showpacs,floatfix,notitlepage,a4paper,twocolumn]{revtex4}
\usepackage{amsfonts,amsmath,units,wasysym,epsfig,graphicx,verbatim,color,subfigure,graphicx,bm,mathrsfs,lipsum,hyperref}
\usepackage[utf8]{inputenc}
\usepackage[normalem]{ulem}  % \sout{old text} for strikeout

\begin{document}

\newcommand{\UNIFE}{Dipartimento di Fisica e Scienze della Terra,  Università di Ferrara, Via Saragat 1, 44122 Ferrara, Italy}
\newcommand{\INFN}{INFN Sezione di Ferrara, Via Saragat 1, 44122 Ferrara, Italy}

\title{Leggett-Garg inequalities and decays of unstable systems}

\author{Francesco Giacosa$^\text{(1,2)}$ and Giuseppe Pagliara$^\text{(3)}$}

\affiliation{$^{\text{1}}$Institute of Physics, Jan Kochanowski University, ul.
Uniwersytecka 7, 25-406 Kielce, Poland,}
\affiliation{$^{\text{2}}$Institute for Theoretical Physics, Goethe University,
Max-von-Laue-Str.\ 1, 60438 Frankfurt am Main, Germany.}
\affiliation{$^3$\UNIFE,\INFN}

\begin{abstract}
We apply the Leggett-Garg inequalities (LGI) to the case of classical and quantum unstable systems. For classical systems the two assumptions of macroscopic realism and non-invasive measurements imply that the three-measurement string $K_3$ is identically equal to one.
Also for quantum mechanical systems --for which the two assumptions are in general not valid-- we find that $K_3=1$ for purely exponential decays ($K_3\leq1$ is the general LGI). On the other hand, the necessary deviations from the exponential decay law at short and long times predicted by quantum mechanics lead to values of $K_3 \neq 1$. Moreover, a strict violations $K_3>1$ occurs typically at short times.  Thus, we conclude that experiments in which such deviations from the exponential decay law have been observed should also find in their data violations of the LGIs.   
\end{abstract}
\maketitle

\section{Introduction}

Correlations between spatially separated entangled states are at the core of Quantum Mechanics (QM) and are a necessary consequence of the linear superposition principle. Such quantum mechanical correlations have no analogous in classical physics and lead to the violations of the Bell's inequalities \cite{Bell:1987hh,Peres:1998sf}. In 1985, Leggett and Garg \cite{Leggett:1985zz} have derived similar inequalities (LGI) for the correlations of the outcomes of measurements of the same observable of a system at different times. 
Interestingly, those violations have been seen in different experimental setups, see the review \cite{review} including also the case of neutrino oscillations \cite{Formaggio:2016cuh}.
 
 The LGI are based on two assumptions which definitely hold true in classical systems: 
 (i)  macroscopic realism (MR), according to which macroscopic properties are uniquely defined, see also \cite{clemente,strict1,strict2,Halliwell:2018dor};
 (ii) non-invasive measurement (NIM), implying that a measurement does not affect in any way the system under investigation. 
 
 A natural question, that we shall address in this work, concerns the violation of the LGI for unstable quantum systems. For such systems, the so-called survival probability $p(t)$ is defined as the the probability that the state has not yet decayed at the time $t>0$, assuming that it was prepared at $t=0$ (thus $p(0)=1$). We recall that an actual decay implies that $p(\infty)=0$, i.e. the Poincaré time is genuinely infinite.

 Quite interestingly, a survival probability can be also defined for strictly classical systems, as for instance the probability that a mouse trap is undecayed, see Fig. 1 for a schematic presentation. 
 In this case, the function $p(t)$ depends on the particular system under study and can have (besides the constraints $p'(t)<0$ and  $p(\infty)=0$) any form. For a classic decay both MR and NIM are clearly fulfilled and, as expected, no violation of the LGI takes place, regardless of the particular classic decay function $p(t)$. In particular, we shall concentrate on the LG correlator $K_3$ which in general fulfills the LGI $-3 \leq K_3\leq 1$. In the case of classic decays, as we will show, it turns out that $K_3=1$, thus the LGI reduces to a LG \textit{equality} in this special case. 
 
For what concerns quantum decays, the survival probability $p(t)$ is usually very well approximated by an exponential function \cite{Weisskopf:1930au,Weisskopf:1930ps}, but it is well established that the exponential behavior is never exact \cite{Fonda:1978dk}. In particular, the deviations are enhanced at short and long times, see also the experimental confirmations in Refs. \cite{raizen1,raizen2,rothe,lastpasca}.
 
At short times, the decay law can be usually described by a quadratic function $p(t)\simeq1-t^2/\tau_Z^2$, where $\tau_Z$ is the Zeno-time. As a consequence, the so-called quantum Zeno effect (QZE), that is the freezing of the decay by subsequent repeated measurements at sufficiently small time intervals, is possible \cite{Degasperis:1973wce,Misra:1976by}. Note, the QZE has been originally verified as a slowing down of certain transitions in systems involving Rabi oscillations between energy levels \cite{Itano:1990zz,balzer,streed}, but it could be later verified for an actual quantum decay in Ref. \cite{raizen2}. 
A related phenomenon is the inverse Zeno effect (IZE): this is an increase of the decay rate that may take place in certain systems when an appropriate time interval between subsequent measurements is chosen. As argued in Refs. \cite{KK1,KK2} the IZE might be as relevant as QZE, see also the theoretical works of Refs.\cite{prlpasca,FPprl,Giacosa:2019nbz,Koshino:2004rw} as well as the experimental verification in \cite{raizen2}. 

In this paper we show, via the correlator $K_3$ mentioned above, that the LGI are violated for quantum decays. In particular, such violations are enhanced when short times are involved, thus when the QZE and/or the IZE are also possible. 
Only in the (unphysical) limit in which the decay is exactly exponential at all times the LGI are not violated and reduces to the LG equality $K_3=1$ that holds for classical decays.

Once verified that the LGI are violated for quantum systems, the natural question is the origin of such deviations. 
Namely, in QM the collapse of the wave function implies necessarily a strong violation of the NIM. Moreover, the MR is also violated, since any quantum decay implies a superposition of decayed/undecayed components: the \textquotedblleft quantum version of the mouse trap\textquotedblright mentioned above, due to the linearity of QM, just as the Schrödinger cat, enters in a corresponding superposition of sprung and not sprung. Yet, as we shall discuss later, the breaking of the NIM (and not of MR) is at the core of the violation of the LGI inequalities for quantum decays. 

The paper is organized as follows: in Sec. II we present the derivation of the LGI for classical and quantum unstable systems; in Sec. III we present some numerical examples that make use of a toy model as well modelling of quantum tunneling as realized in experiments; finally, in Sec. IV, we present our conclusions.

\begin{figure}
\centering
\epsfig{file=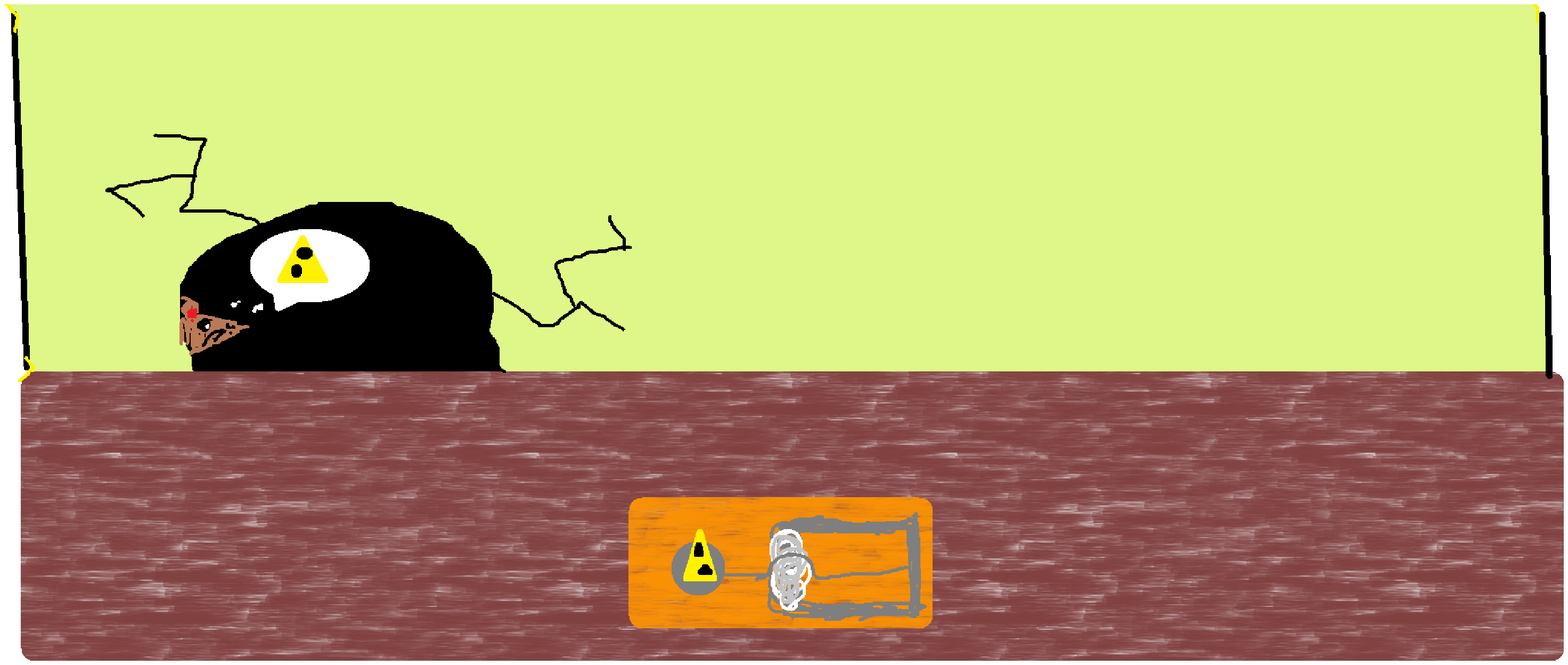,height=5cm,width=7cm,angle=0}
\vskip 0.2cm
\centering
\epsfig{file=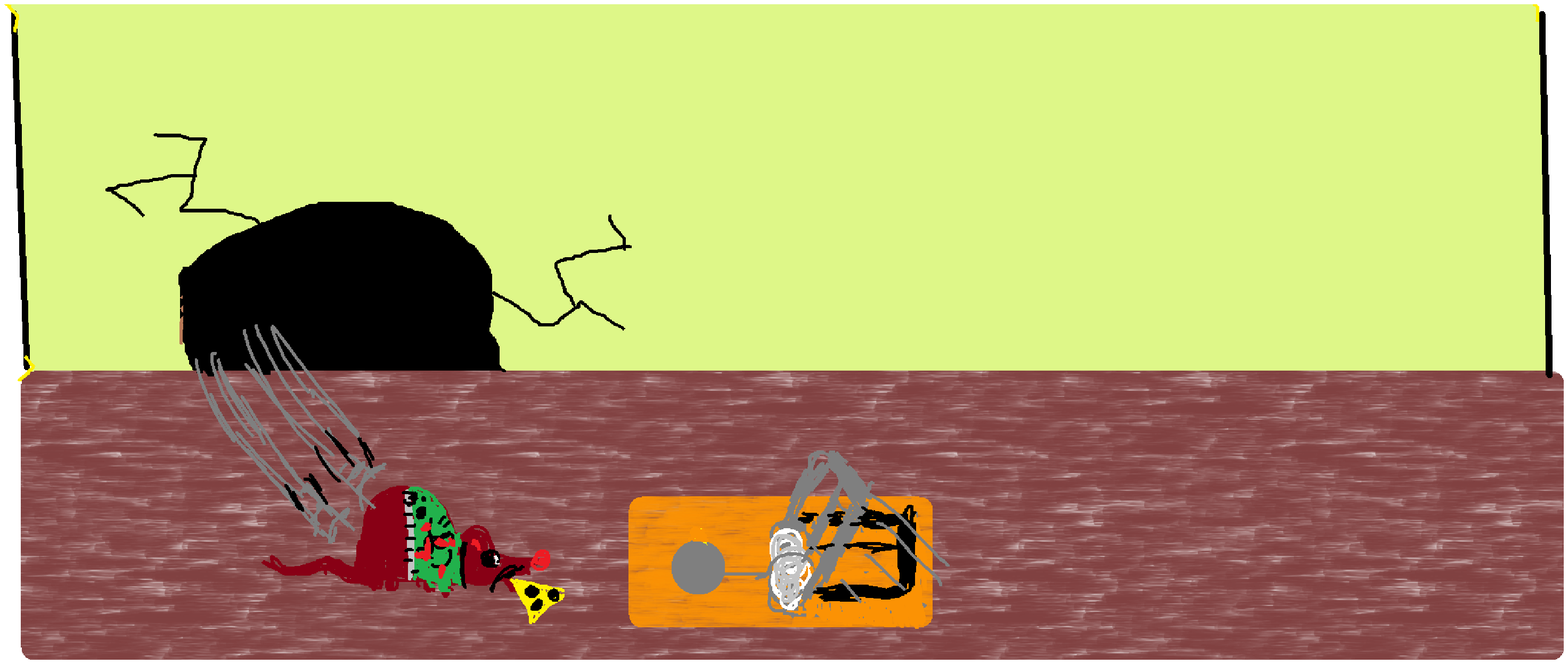,height=5cm,width=7cm,angle=0}
\vskip 0.2cm
\centering
\epsfig{file=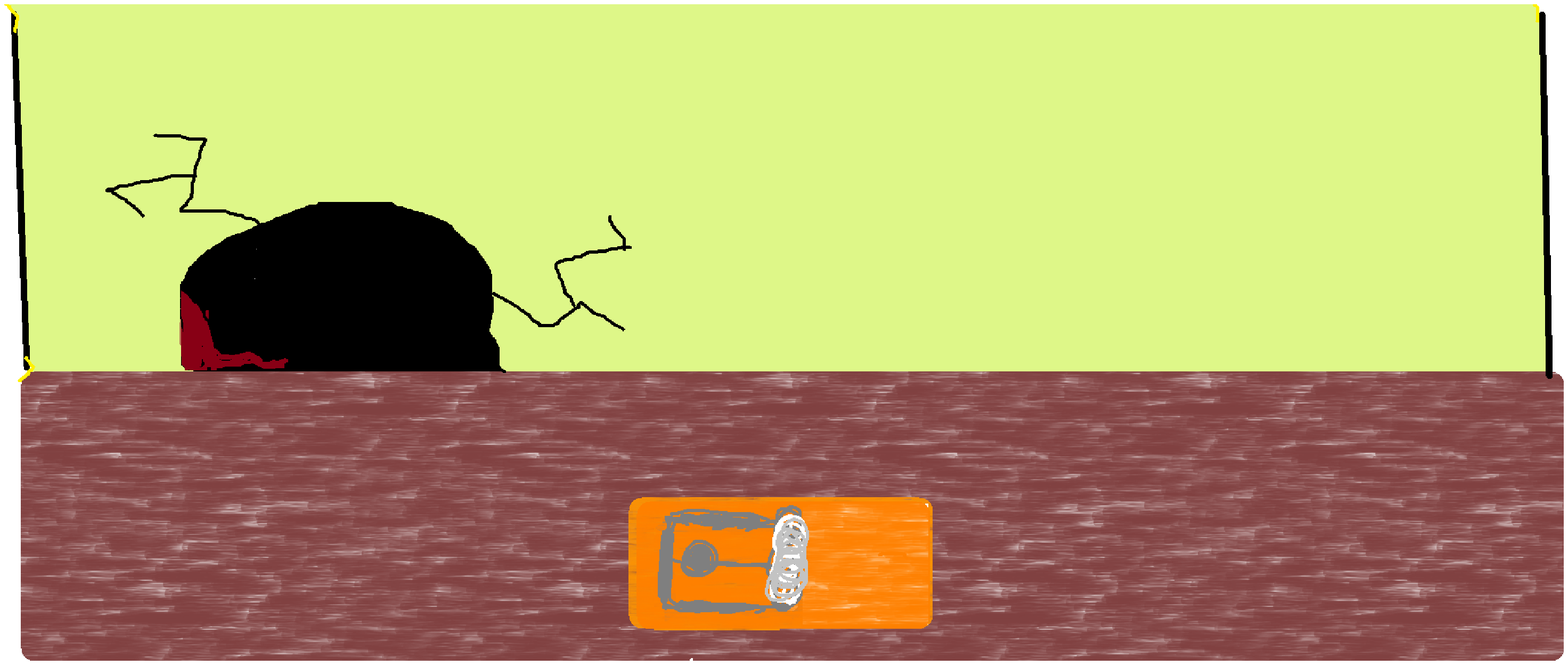,height=5cm,width=7cm,angle=0}
%\includegraphics[scale=0.29]{TOPO.pdf}
%\centering
%\includegraphics[scale=0.29]{TOPO2.pdf}
%\centering
%\includegraphics[scale=0.29]{TOPO3.pdf}
\label{fig:1}
\caption{A classical decay via a mouse trap. Upper panel: a trap is placed at the initial time $t=0$, the mouse notices it; the trap is undecayed (U). Middle panel: at a certain time $0<t_0<\infty$ the mouse very rapidly snaps the cheese; the traps decays at this time. Lower panel: the trap is decayed (D) for any time $t>t_0$. }
\end{figure}

\vskip 1cm
\begin{figure*}[!ht]
	\begin{centering}
		\epsfig{file=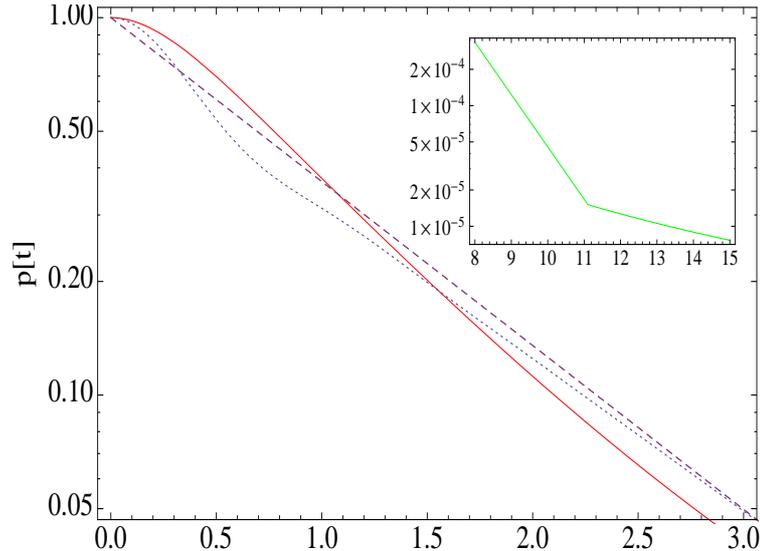,height=8cm,width=10cm,angle=0}
		\caption{The three different non-exponential functions used for $p(t)$. The red continuous line corresponds to Eq. \ref{p1}, the blue dotted to Eq. (\ref{raizen}) (the one 
	  reported in Ref. \cite{Reizen1})	and 
		the green (light gray) line in the insert to Eq. (\ref{rothe}), see Ref. \cite{rothe}. The dashed line is the exponential function.}
	\label{prima}
	\end{centering}
\end{figure*}

\section{Leggett-Garg inequalities for unstable systems}

The starting point of the LGI is the n-measurement LG string $K_n$ which is built from the two-times correlation functions
$C_{ij}$ which read:
\begin{equation}
    C_{ij}=\sum_{Q_i,Q_j}Q_iQ_jP_{ij}(Q_i,Q_j) \text{ ,}
\end{equation}
where $Q_i (Q_j)$ represents the outcome of a measurement at the time $t_i (t_j)$
with $t_j>t_i$, which we set to $1$ if the system is found to be still undecayed (or \textquotedblleft alive\textquotedblright) and $-1$ if it is found to be decayed (or \textquotedblleft dead\textquotedblright). 
The quantity $P(Q_i,Q_j)$ is the joint probability associated to the four possible events $Q_i=\pm1$ and $Q_j=\pm1$.  For instance, $P_{ij}(1,1)$, is the joint probability that after the first and the second measurement the system has been found undecayed (and similar for the other three joint probabilities).
Thus, in the study of (both classic and quantum) decays the correlation quantity $C_{ij}$ takes the explicit form
\begin{equation}
    C_{ij}=P_{ij}(1,1)+P_{ij}(-1,-1)-P_{ij}(1,-1)-P_{ij}(-1,1) \text{.}
    \nonumber
\end{equation}
In the following, we shall concentrate our discussion on the LG string $K_3$ given by \cite{review}
\begin{equation}
 K_3=C_{12}+C_{23}-C_{13} \text{ ,}   
 \label{k3}
\end{equation} 
which  is constrained to fulfill the LGI:
\begin{equation}
-3\leq K_3\leq 1 \text{ .}
\end{equation}

\subsection{Classical case}

Let us discuss now how to compute the correlation function
for a \textquotedblleft classical unstable system\textquotedblright, namely a system and a measurement of it that obey the MR and the NIM assumptions. 
Within this context, it is easy to realize that $P_{ij}(-1,1)=0$ since, if the system is decayed a $t_i$, it cannot be alive at $t_j$. This holds true in all cases (classical or quantum) in which the measurements are sequential since the second choice of the
measurement, does not influence the first outcome.

Then, we can set $P_{ij}(-1,-1)$ equal to $P_i(-1)$, which is the probability that the system has decayed at the  time $t_i$. Namely, if the system decayed at $t_i$ then it is surely still decayed at $t_j>t_i$ and thus the joint probability corresponds to the single measurement probability. 

Alternatively, one can write: 
\begin{equation}
 P_{ij}(-1,1) + P_{ij}(-1,-1) = P_i(-1) \text{ ,}
\end{equation}
which corresponds to summing over the two possible states at $t=t_j$ and, since $P_{ij}(-1,1)=0$, one obtains:
\begin{equation}
P_{ij}(-1,-1)=P_i(-1) \text{ .} 
\end{equation}
Notice that this is strictly true under the MR hypothesis since the state of the particle is decayed or undecayed regardless of the measurement.
Similarly:
\begin{equation}
P_{i,j}(1,1)=P_j(1) \text{ ,} 
\end{equation}
which is a direct consequence of of the NIM hypothesis. In fact, it corresponds to the assumption that the mouse is \textit{not} affected by the measurement that has occurred at $t_i<t_j$. We will see that in the quantum case this joint probability is different, since the NIM is not fulfilled. 

The joint probability $P_{ij}(1,-1)$ can be obtained by the normalization  $\sum_{Q_i,Q_j}P_{ij}(Q_i,Q_j)=1$,
thus $P_{ij}(1,-1)=1-P_j(1)-P_i(-1)$. 
Finally, from $P_i(1)+P_i(-1)=1$ it follows that
\begin{equation}
C_{ij}=1+2P_j(1)-2P_i(1) 
\label{ccl}
\end{equation}

It is also useful to re-express the previous equation by introducing the classic survival probability $p(t)$ that the system has not decayed at the time $t$. One has: 
\begin{equation}
P_j(1)=p(t_j) \text{,}\,\, P_j(-1)=1-p(t_j) \text{ ,}
\end{equation}
out of which the quantity $C_{ij}$ takes the form: 
\begin{equation}
C_{ij}=1+2p(t_j)-2p(t_i) \text{ .}
\label{ccl2}
\end{equation}
The summary of all classic probabilities is displayed in Table I, where a comparison with the QM case (to be discussed later on) can be found. 

Out of Eq. (\ref{k3}) we obtain via a straightforward calculation:  
\begin{equation}
K_3=1 \text { ,}
\end{equation} 
that holds for \textit{each} classical unstable system.
This is a quite remarkable result since it does not even depend on the specific 
functional form of the classical decay law $p(t)$, which could be very well different from an exponential.

The classical decay can be explained with a simple example. Following a certain established tradition for QM related topics, we pick up an animal, a mouse. In a given room adjacent to the mouse lair, an old fashioned mouse trap with cheese is placed at $t=0$. The mouse is associated to a certain probability $p(t)$ that it has not yet got in contact with the trap. Of course, $p(t)$ is a given function related to the complicated and stochastic algorithm of the mouse brain and is not known a priori. We simply assume that $p(t)$ tends to zero for large times, hence at a certain (unknown) time the mouse will steel the cheese, see Fig. 1 for a pictorial representation of the mouse-trap sequence.
Note, in the spirit of the time and according to the animal friendly attitude of the authors, we assume that our mouse -even if only ideal- is not injured in the process: it takes the cheese and runs away content. Yet, by doing so the trap has sprung. 
The observer (within the saga an unpleasant old-fashioned farmer dealing with old mouse traps) opens the room at the time $t_i>0$ and $t_j>t_i$ to see if the mouse was there: (s)he checks if the trap is still undecayed (U) or decayed (D) at both times and studies the correlation $C_{ij}$.  Clearly, if the trap has sprung at $t_i$ is also such at $t_j$: this is the sequence DD. Vice-versa, if it is not sprung at $t_j$, it was not such also at $t_i$ (sequence UU). Since DU is zero, the last sequence is UD: the trap is intact at  $t_i$ but sprung at $t_j$. The probability for UD
is calculated as the probability that the system decays between $t_i$ and $t_j$.  By denoting with $h(t)=-p'(t)$ the probability density of decaying at the time $t$, UD is given by
\begin{equation}
\int_{t_i}^{t_j}h(t)\mathrm{d}t = p(t_i)-p(t_j) \text { .}   
\end{equation}
Summarizing (see also Table I):
\begin{eqnarray}
C_{ij} &=&\text{UU}+\text{DD}-\text{UD}  \nonumber \\
&=&p(t_j)+\left[1-p(t_{i})\right] -\left[ p(t_{i})-p(t_{j})\right] 
\text{ ,}
\end{eqnarray}
in agreement with Eq. (\ref{ccl}), out of which $K_3$ is easily evaluated to be $1$, independently on the `mouse function' $p(t)$. 

\begin{center}
\begin{tabular}{|l|l|l|}
\hline
Sequence & Classic (MR+NIM) & QM (collapse) \\ \hline
$P_{ij}(1,1)\equiv $UU & $p(t_{j})$ & $p(t_{i})p(t_{j}-t_{i})$ \\ \hline
$P_{ij}(1,-1)\equiv $UD & $p(t_{i})-p(t_{j})$ & $p(t_{i})\left[
1-p(t_{j}-t_{i})\right] $ \\ \hline
$P_{ij}(-1,1)\equiv $DU & $0$ & $0$ \\ \hline
$P_{ij}(-1,-1)\equiv $DD & $1-p(t_{i})$ & $1-p(t_{i})$ \\ \hline
Sum & $1$ & $1$ \\ \hline
\end{tabular}
Table I: Probabilities for two measurements at $t_{i}$ and $t_{j}$ (U = undecayed, D = decayed). 
\end{center}

\vskip 1cm
\begin{figure*}[!ht]
	\begin{centering}
		\epsfig{file=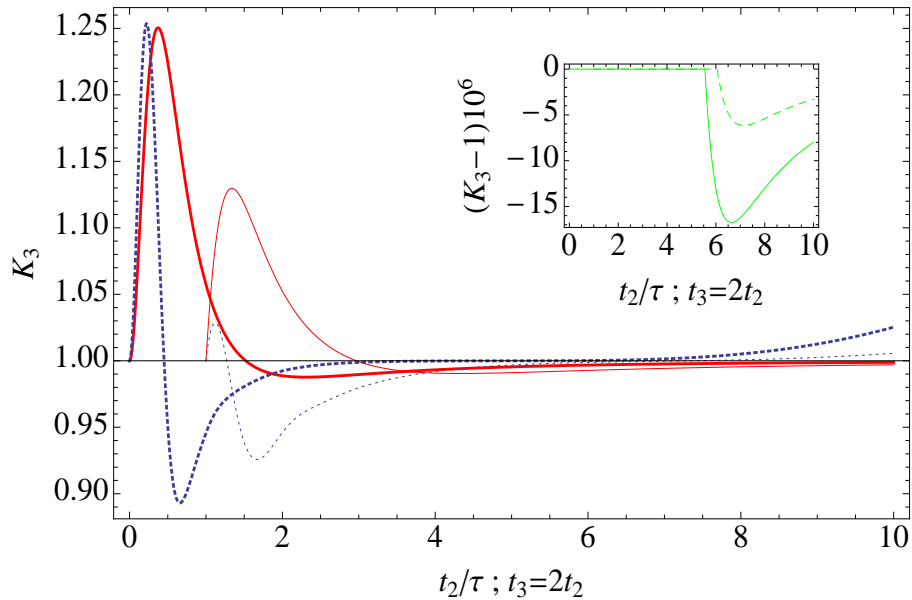,height=8cm,width=12cm,angle=0}
		\caption{$K_3$ as a function of $t_2$ and with $t_3=2t_2$ for the three different $p(t)$. The solid red lines refer to $t_1=0$ (thick) and $t_1=\tau$ (thin) for the $p(t)$ of eq.20.
		The dotted blue lines refer to $t_1=0$ (thick) and $t_1=\tau$ (thin) for the $p(t)$ of eq.21.
		The green (light gray) lines in the insert show the the difference $K_3-1$ (magnified by a factor of $10^6$) for the long time deviations case of eq.22 for $t_1=0$ (solid line) and $t_1=\tau$ (dashed line). }
	\label{seconda}
	\end{centering}
\end{figure*}

\subsection{Quantum systems}
Let us now discuss the case of an unstable quantum system. Both MR and NIM are violated, thus 
the system is, if not observed, in a superposition of undecayed and decayed configurations (no MR); moreover the act of observing/measuring the system perturbs its decay law by resetting the clock (no NIM).

Yet, even if MR cannot be assumed to hold, each measurement generates a collpase of the system into either decayed or not decayed. In this respect, the collapse is equivalent to MR, since it is not possible -within the present setup- to distinguish MR from the collapse. Note, the collapse is intended here as an effective phenomenon, whose deep understanding is still not achieved (it is not clear if it is a physical collapse or not, e.g. Ref. \cite{Bassi:2012bg}). Yet, as matter of fact, for each observer the outcome of the measurement is univocal, either decayed or not, and this is enough for the following discussion. In other words, in the study of decays we need to work with the decayed/undecayed basis and we cannot \textit{rotate} to another basis to test the QM superposition.

\begin{figure}[!ht]
\begin{centering}
\centering
				\epsfig{file=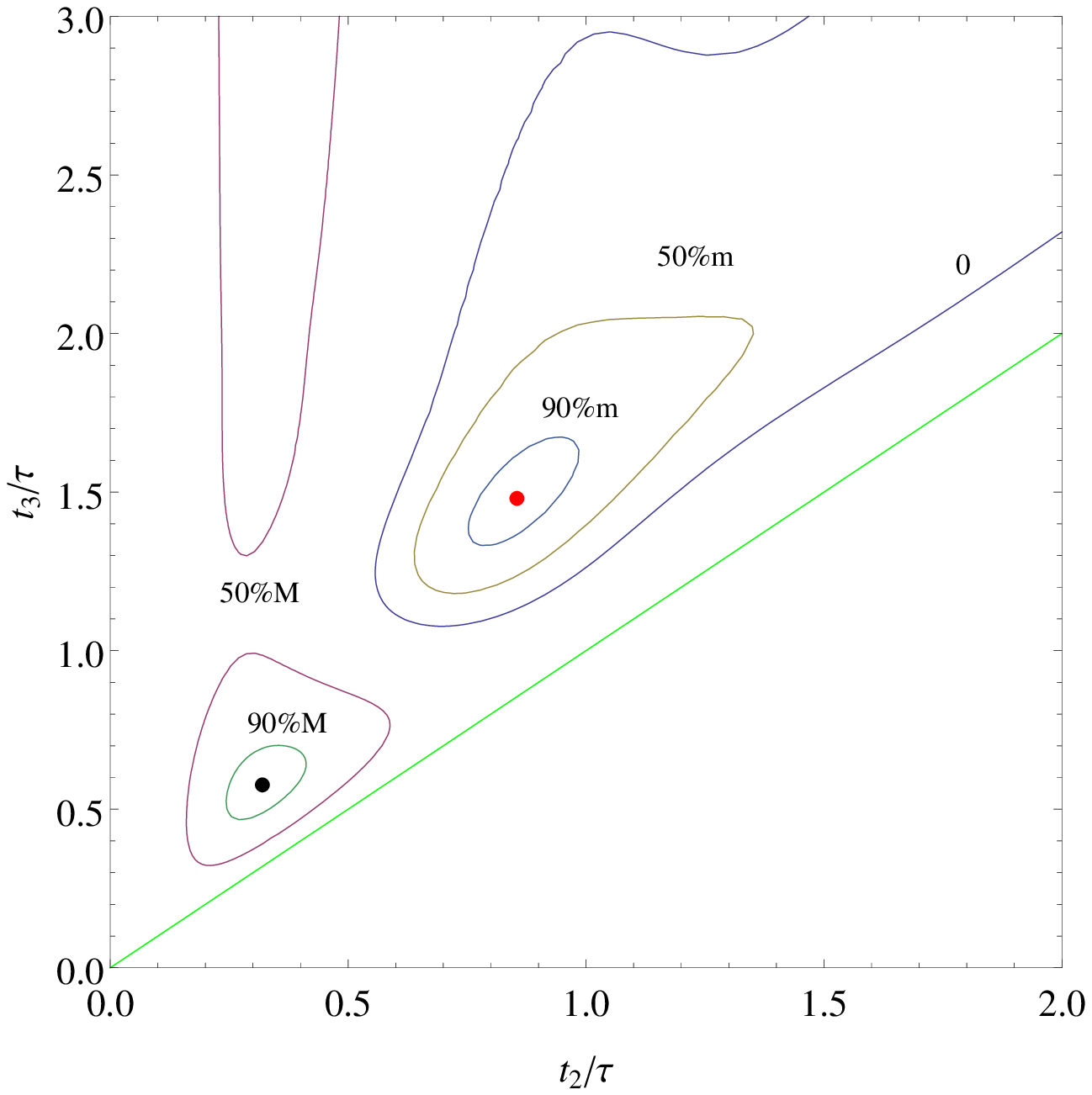,height=8cm,width=8cm,angle=0}
			
\vskip 1.4cm

\centering
			\epsfig{file=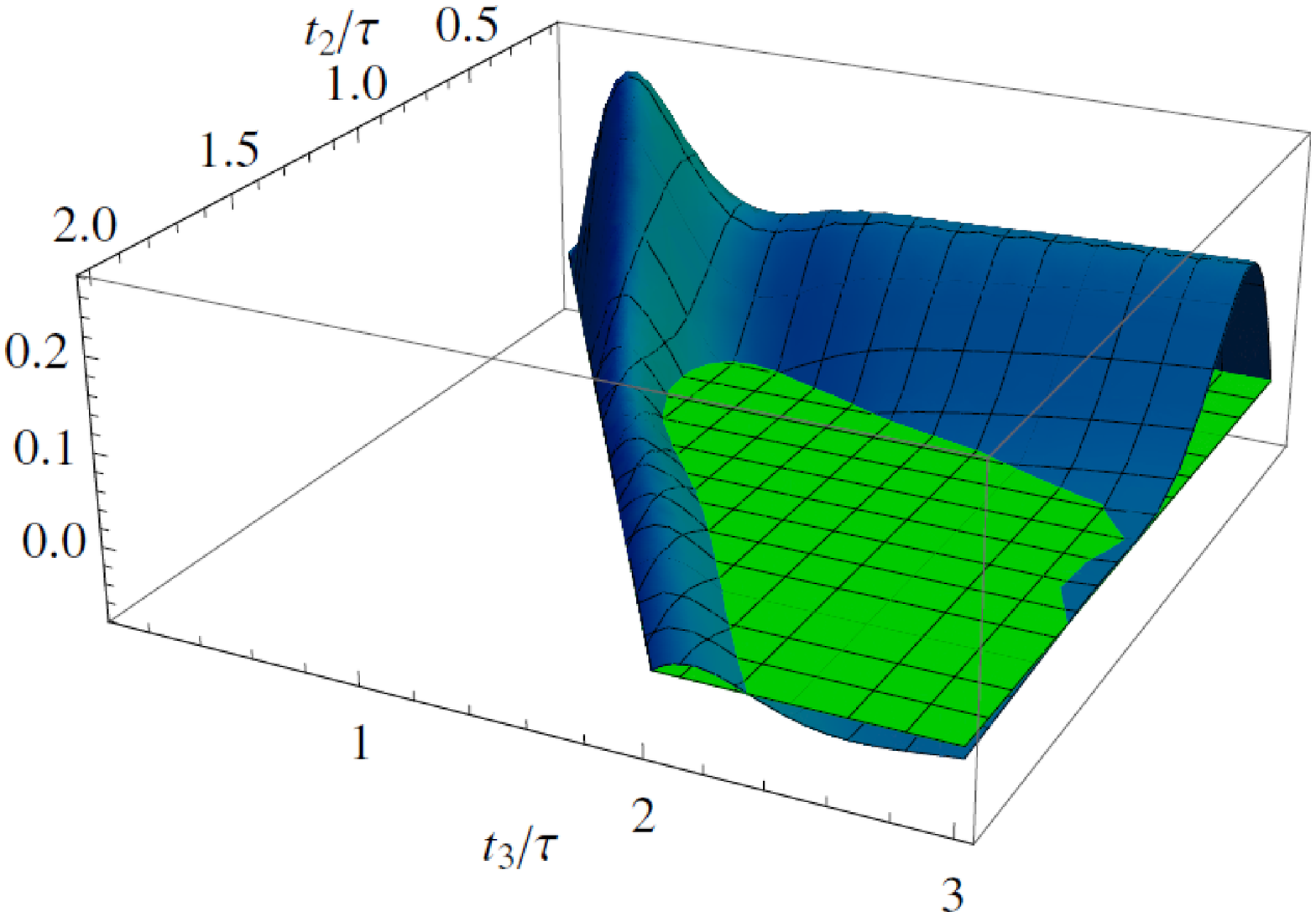,height=8cm,width=8cm,angle=0}

				\label{fig:4}
	
			\caption{Upper panel: Contours of $K_3-1$ for $t_1=0.1 \tau$ for the $p(t)$ of Eq.9. The dots correspond to the maximum (M=0.28 for this example, black) and the minimum (m=-0.08 for this example, red). The green line delimits the region of the plane for which $t_3>t_2$. The LGI is violated and oscillations around zero are visible. Lower panel: 3D plot of $K_3-1$ (blue surface) corresponding to the upper panel. In green, the plane $K_3-1=0$. }
	\end{centering}
\end{figure}

Next, we turn to the evaluation of the the three joint probabilities ($P_{ij}(-1,1)=0$ as before, since D is once for all).
It is useful to introduce the conditional probability 
$P(jQ_j|iQ_i)$ which is the probability of obtaining $Q_j$ provided that at $t_i$ the system had a value $Q_i$. Through the conditional probability one can write:
$P_{ij}(1,1)=P(j1|i1) P_i(1)$. Now, if the system was alive at $t_i$, the probability that it is still alive at $t_j$  is $p(t_j-t_i)$, 
since the system has collapsed onto the undecayed state after the first measurement. 
This is the crucial difference between the classical and the quantum cases: the measurement \textquotedblleft resets\textquotedblright the clock to the initial time; this process is a clear violation of NIM \footnote{We assume that the quantum system is not subject to the interaction with
the environment, that is, it is not subject to decoherence. Namely, decoherence
would imply an effective measurement of the state of the system performed by the environment. If that is the case, one should include this additional interaction explicitly into account.}. The same features of QM is at the origin of the QZE and IZE.
We can thus write 
\begin{equation}
P_{ij}(1,1)=p(t_i)p(t_j-t_i).
\end{equation}
Let us compare this expression with the same joint probability in the classical case, which has been previously derived as $P_{ij}(1,1)=P_j(1)=p(t_j)$. That result can be re-obtained by considering that in the classical case (MR and NIM hold true) the conditional probability $P(j1|i1)=P_j(1)/P_i(1)$ since the condition that the system was undecayed at $t_i$ is necessary for it to be undecayed at $t_j$ (namely, the set of cases in which the system is undecayed at $t_j$ is a subset of the set of cases in which the system is undecayed at $t_i$). Note, the quantum mechanical $P(j1|i1)=p(t_j-t_i)$ reduces to the classical one in the case of a purely exponential decay law.

Next, similarly to the classical case, $P_{ij}(-1,-1)=P_i(-1)=1-p(t_i)$.
As before $P_{ij}(1,-1)$ can be determined by the normalization condition and finally the quantum two-times correlation function reads:
\begin{equation}
    C_{ij}^q=1+2p(t_i)p(t_j-t_i)-2p(t_i)
    \text{ ,}
    \label{cq}
\end{equation}
compare with Eqs. (\ref{ccl}) and (\ref{ccl2}). 
If the decay law of a quantum mechanical unstable system was purely exponential, $p(t)=e^{-\gamma t}$, then one would obtain a result in agreement with Eq. (\ref{ccl2}). 
We summarize the quantum results and compare them to the classical outcomes in Table I. 

In QM, the quantity $K_{3}$ takes the form 
\begin{eqnarray}
K_{3} &\equiv &K_{3}(t_{1},t_{2},t_{3})=1+2p(t_{1})\left[
p(t_{2}-t_{1})-p(t_{3}-t_{1})\right]   \nonumber \\
&&+2p(t_{2})p(t_{3}-t_{2})-2p(t_{2}) \text{ .} 
\end{eqnarray}

A specific choice, useful later, is obtained by setting $t_{1}=0$: 
\begin{eqnarray}
K_{3}(0,t_{2},t_{3}) &=&1+2\left[ p(t_{2})-p(t_{3})\right]   \nonumber \\
&&+2p(t_{2})p(t_{3}-t_{2})-2p(t_{2})\text{ .}
\end{eqnarray}

For $p(t)=e^{-\gamma t}$ one gets $K_{3}(t_1,t_2,t_3)=1$.  
This is consistent with the fact that the exponential decay carries no memory.
Yet, as discussed in the introduction, the actual QM decay law is never exactly exponential, even if the exponential law can be a very good approximation. For a general discussion, let us consider the
following simplified schematic form for $p(t)$ \cite{factorzeta,lastpasca}:
\begin{equation}
p(t)\simeq\left\{ 
\begin{array}{c}
1-\frac{t^{2}}{\tau _{Z}^{2}}\text{ for small }t \\ 
Ze^{-\gamma t\text{ }}\text{ for intermediate }t \\ 
kt^{-\alpha }\text{ for large }t%
\end{array}%
\right.
\label{schem}
\end{equation}%
where $\tau _{Z}$ (Zeno-time), $Z$, and $k$ are appropriate factors whose
numerical values depend on the specific system under study. 

The short-time deviations, already discussed in the introduction, allow for QZE. At intermediate times, the behavior is exponential, but a constant $Z$ different from one enters into the expression. 
It can be both larger or smaller than one, depending on the particular quantum decay. Indeed, $Z>1$
implies QZE and $Z<1$ is a manifestation of IZE. [In fact, within the exponential regime, a single measurement at $T$ gives
the survival probability $Ze^{-\gamma T},$ while two measurements at $T/2$
and $T$ respectively correspond to $Z^{2}e^{-\gamma T},$ thus  one has QZE
for $Z>1$ and IZE for $Z<1$]. 
At long times, as shown already in the seminal paper of Ref. \cite{khalfin}, the
function $p(t)$ shows a power law behavior, the reason of which is the necessary existence of a ground state: the decay law at large $t$ is determined by the behavior of the spectral function at energies close to the ground state energy. Typically, the power law behavior occurs after many life times, order of 10 , see Ref. \cite{rothe} where such a challenging
measurement has been performed.

If we choose all three times $t_{1},t_{2},t_{3}$ within the intermediate
\textquotedblleft exponential\textquotedblright\ region, we find%
\begin{equation}
K_{3}\equiv K_{3}(t_{1},t_{2},t_{3})\simeq 1+2Z(Z-1)e^{-\gamma t_{2}}\neq 1
\label{K31}
\end{equation}%
which depends only on the intermediate times $t_{2}.$
It implies that $K_{3}>1$ when $Z>1$ and vice-versa. Anyway, in both cases one obtains a result different from the classic LG result $K_{3}=1.$ The
classic result is obtained only for $t_{2}$ large enough (but still within the exponential interval).

If, instead, we choose $t_{1}=0$ and $t_{2},t_{3}$ within the exponential
domain, we get a quite analogous result:  
\begin{equation}
K_{3}\equiv K_{3}(0,t_{2},t_{3})\simeq 1+2Z(Z-1)e^{-\gamma t_{3}} \text{ ,}
\label{K32}
\end{equation}%
where, in this case, the final-time $t_{3}$ enters into the expression.

Before showing in the next subsection some numerical results for some specific
models, we briefly discuss the origin of the LGI violations: while a quantum decay violates both the MR and the NIM, it is only the latter that is relevant for the LGI violation. This feature is already clear from our discussion above about the f.a.p.p. equivalence between the classic MR and the collapse in QM. In addition, it can be also understood by the following arguments:

(i) In the purely exponential limit, the quantum decay does not break the LGI. In this particular case, there is no difference between an invasive and a non-invasive measurement (see Table I): the NIM de facto applies (even if a collapse takes place when a measurement is performed, but the reset of the clock is invisible in the exponential limit). Yet, the quantum state (alias the quantum version of the mouse trap) is in a superposition of decayed and undecayed and thus breaks MR. The non-violation of the LGI in this case implies that the violation of MR alone is not sufficient. 

(ii) Conversely, let us consider the classical example of the mouse in which, however, the mouse is affected by the observer checking the status of the trap at a given time. For instance, the mouse may reset its own internal clock when someone opens the room by looking at the trap. Then, it is clear that the NIM is broken in this classical example, but MR is not, since the trap is always in one unique state, either decayed or not. The resulting equations are the same as in the quantum case described above and the LGI are violated. Thus, this example shows that the  violation of MR is also not necessary for violating LGI.

Both (i) and (ii) show that the NIM alone is responsible for the breaking of the LGI in the case under study. The important aspect is that in QM the NIM is never fulfilled (a part from the limiting case of an exponential), while in a classical system this can be in principle always realized (in a classic word, we can always find a way to check if the trap is sprung or not \textit{without} the mouse noticing it). 

\section{Numerical examples on non-exponential decay laws}
 In this section we describe some specific numerical examples. 
Let us first introduce a toy decay law which features both the short and the long time deviations from the exponential:
\begin{equation}
p(t)=\frac{1}{2}\left( e^{-\gamma \frac{t^{2}}{t+1}}+\frac{1}{1+t^{\alpha }}%
\right) \text{ .}  \label{p1}
\end{equation}
The corresponding temporal behaviour is shown in Fig. 2 (red solid line) for $\gamma=1$ and $\alpha=2$.
For comparing with the exponential decay law, we have fitted $p(t)$ with an exponential and found an effective life-time $\tau$ (that we use as unit of time). 

%------------------------------------------------

Next, let us consider a more realistic, and hence interesting, decay law which has been found in the experimental setup of cold atoms
devised in Refs. \cite{Reizen1,raizen2}.
In particular, we make use of the analytical approximation of the tunneling process used in those works:
\begin{eqnarray}
\log(p(t))&=&-\int_0^t \mathrm{d}\tau (t-\tau)W(\tau) \text{ ,} 
\nonumber
\\
W(\tau)&=&\frac{a^2}{2V_0}\int_{-\infty}^{\infty}\mathrm{d}s\frac{1}{1+(s-a\tau/V_0)^2}\frac{1}{1+s^2}\nonumber \\
&&\cos\left(\frac{V_0^2}{a}\int_{s-a\tau/V_0}^s \sqrt{1+z^2}\mathrm{d}z\right) \text{ .}
\label{raizen}
\end{eqnarray}

This $p(t)$ depends on two parameters, the acceleration of the trap $a$ and the potential well depth $V_0$. For showing a numerical example we fix $V_0=100$ kHz/h and $a=4200$m s$^{-2}$ and the time dependence of the survival probability is shown in Fig. \ref{prima}, blue dotted line. The clearly visible deviations from the exponential at short times are in agreement with the experimental findings of Refs. \cite{Reizen1,raizen2}.

For completeness, we also consider an example long times deviations from the exponential. We use the results of Ref. \cite{rothe} for the decays of molecules of polyfluorene ($\tau=0.35$ ns, power law index $\alpha=-2.26$ and turnover time $\tau^{\mathrm{turnover}}=11.1\tau$), whose survival probability can be modelled as:

\begin{equation}
p(t)=\left\{
\begin{array}
[c]{c}%
e^{-t/\tau}\text{ for }t\leq\tau^{\text{turnover}}\\
kt^{\alpha}\text{ for }t>\tau^{\text{turnover}}%
\end{array}
\right.  
\text{ ,}
\label{rothe}%
\end{equation}
where k is a normalization constant. (Note, in comparison with Eq. (\ref{schem}) we neglect the initial quadratic time and set $Z=1$.) The deviations from the classical result are displayed in the insert of Fig. \ref{prima}, green (light gray) line.

It is interesting to observe that, in principle, the QZE/IZE is also possible for measurements performed at very long time intervals that span into the power-law behavior. However, one would need to detect the same unstable system at least twice and find it undecayed in both cases. This is a quite improbable event that would require a very large statistic that is not reachable at present.

In Fig. \ref{seconda} we show $K_3$ as a function of $t_2$ for
two values of $t_1=0,\tau$ and for $t_3=2t_2$.
In all three cases deviations from the classical limit$K_3=1$ are found: only from above for the first example of $p(t)$, from above and below  in the case of the $p(t)$ of Ref. \cite{Reizen1}, an only from below for the case of long time deviations.
Strictly speaking, the general LGI $K_3\leq1$, is violated by the first and the second $p(t)$, but all of them violate the LG equality $K_3=1$ that holds for classic decays. 
The magnitude of the departures from $K_3=1$ are quite different: they amount to  $10\%$ or more for \cite{Reizen1}, while they are very small, order of $10^{-5}$,
for the case of long time deviations of Ref. \cite{rothe}.
Moreover, we observe that the violations are of the order of a few $\%$ even in the case in which $t_1 \sim \tau$, see Fig.3. This property may be interesting to investigate non-exponential decays and QZE and IZE also when studying other unstable systems. 

For instance, as computed in Ref. \cite{FPatoms} for the electromagnetic transition of the hydrogen atoms, $\tau_Z \sim 10^{-15}$ sec while $\tau \sim 10^{-9}$ sec, thus a direct experimental detection of QZE and IZE would be very challenging. On the other hand our results suggest that the correlations functions built for testing the LGI could show sizable and potentially detectable deviations from the classical/exponential case. A viable possibility would be also to investigate the functions $K_3(t_1,t_2,t_3)$ and $K_3(0,t_2,t_3$, that are different form unity even when the times belong to the exponential domain, see Eqs. (\ref{K31}) and (\ref{K32}), respectively. The study of the correlator $K_3$ allows to investigate at the same time the deviations from the exponential decay as well as QZE and IZE. Moreover, it does so by using only two or three intermediate measurements, what can be a simplification in actual future realizations.  

As a final example, we also display the contour plots of $K_3(0.1\tau,t_2,t_3)-1$ for the survival probability function of Eq. (\ref{raizen}) with $t_1=0.1 \tau$ as well as the corresponding 3D plot, see Fig. 4. In this way the landscape of departures from $K_3=1$ are visible.

\section{Conclusions}

In this work, we have studied the LGI for classic and quantum decays by focusing on the three-times correlator $K_3$. The latter equals unity for any classic decay. In the quantum case, it is such \textit{only} in the (unphysical) limit in which  the decay law is purely exponential, but is different as soon as deviations are taken into account. Since the quantum decay law is never purely exponential but displays deviations at short and long times, $K_3 \neq 1$. Interestingly, such violations are enhanced at short times and are connected to the QZE and IZE, but are also present at long times.

We have provided numerical examples of such violations of the LGI also by using data from decays already measured in experiments aiming at testing the short and the long time behavior of the quantum decay law.
The study presented in this work offers an additional tool to test the non-exponential behaviour of the quantum decay law by measuring correlation functions. In particular, detecting such violations could be easier than detecting the QZE and IZE within the initial quadratic regime since the departures from $K_3=1$ may last longer. 

In the future, the study of LGI can be applied to various systems, such as the one described in Ref. \cite{lastpasca}. Moreover, due to the ability to model potentials, it can be applied to novel tunneling experiments. Another interesting extension concerns the case of multiple decays \cite{Giacosa:2019jxz,Giacosa:2011xa,Giacosa:2021hgl}, in which more than a single decay channel is considered.

\vskip 0.5cm
\textbf{Acknowledgments:} 
The authors thank Angela Giacosa for preparing figure 1.

%\bibliography{references}

\begin{thebibliography}{35}
\expandafter\ifx\csname natexlab\endcsname\relax\def\natexlab#1{#1}\fi
\expandafter\ifx\csname bibnamefont\endcsname\relax
  \def\bibnamefont#1{#1}\fi
\expandafter\ifx\csname bibfnamefont\endcsname\relax
  \def\bibfnamefont#1{#1}\fi
\expandafter\ifx\csname citenamefont\endcsname\relax
  \def\citenamefont#1{#1}\fi
\expandafter\ifx\csname url\endcsname\relax
  \def\url#1{\texttt{#1}}\fi
\expandafter\ifx\csname urlprefix\endcsname\relax\def\urlprefix{URL }\fi
\providecommand{\bibinfo}[2]{#2}
\providecommand{\eprint}[2][]{\url{#2}}

\bibitem[{\citenamefont{Bell}(1987)}]{Bell:1987hh}
\bibinfo{author}{\bibfnamefont{J.~S.} \bibnamefont{Bell}},
  \emph{\bibinfo{title}{{Speakable and Unspeakable in Quantum Mechanics.
  Collected papers on Quantum Philosophy, Cambridge, UK: Univ. Pr.}}}
  (\bibinfo{year}{1987}).

\bibitem[{\citenamefont{Peres}(1999)}]{Peres:1998sf}
\bibinfo{author}{\bibfnamefont{A.}~\bibnamefont{Peres}},
  \bibinfo{journal}{Found. Phys.} \textbf{\bibinfo{volume}{29}},
  \bibinfo{pages}{589} (\bibinfo{year}{1999}), \eprint{quant-ph/9807017}.

\bibitem[{\citenamefont{Leggett and Garg}(1985)}]{Leggett:1985zz}
\bibinfo{author}{\bibfnamefont{A.}~\bibnamefont{Leggett}} \bibnamefont{and}
  \bibinfo{author}{\bibfnamefont{A.}~\bibnamefont{Garg}},
  \bibinfo{journal}{Phys. Rev. Lett.} \textbf{\bibinfo{volume}{54}},
  \bibinfo{pages}{857} (\bibinfo{year}{1985}).

\bibitem[{\citenamefont{Emary et~al.}(2013)\citenamefont{Emary, Lambert, and
  Nori}}]{review}
\bibinfo{author}{\bibfnamefont{C.}~\bibnamefont{Emary}},
  \bibinfo{author}{\bibfnamefont{N.}~\bibnamefont{Lambert}}, \bibnamefont{and}
  \bibinfo{author}{\bibfnamefont{F.}~\bibnamefont{Nori}},
  \bibinfo{journal}{Reports on Progress in Physics}
  \textbf{\bibinfo{volume}{77}}, \bibinfo{pages}{016001}
  (\bibinfo{year}{2013}).

\bibitem[{\citenamefont{Formaggio et~al.}(2016)\citenamefont{Formaggio, Kaiser,
  Murskyj, and Weiss}}]{Formaggio:2016cuh}
\bibinfo{author}{\bibfnamefont{J.}~\bibnamefont{Formaggio}},
  \bibinfo{author}{\bibfnamefont{D.}~\bibnamefont{Kaiser}},
  \bibinfo{author}{\bibfnamefont{M.}~\bibnamefont{Murskyj}}, \bibnamefont{and}
  \bibinfo{author}{\bibfnamefont{T.}~\bibnamefont{Weiss}},
  \bibinfo{journal}{Phys. Rev. Lett.} \textbf{\bibinfo{volume}{117}},
  \bibinfo{pages}{050402} (\bibinfo{year}{2016}), \eprint{1602.00041}.

\bibitem[{\citenamefont{{Clemente} and {Kofler}}(2015)}]{clemente}
\bibinfo{author}{\bibfnamefont{L.}~\bibnamefont{{Clemente}}} \bibnamefont{and}
  \bibinfo{author}{\bibfnamefont{J.}~\bibnamefont{{Kofler}}},
  \bibinfo{journal}{\pra} \textbf{\bibinfo{volume}{91}}, \bibinfo{eid}{062103}
  (\bibinfo{year}{2015}), \eprint{1501.07517}.

\bibitem[{\citenamefont{{Knee} et~al.}(2016)\citenamefont{{Knee}, {Kakuyanagi},
  {Yeh}, {Matsuzaki}, {Toida}, {Yamaguchi}, {Saito}, {Leggett}, and
  {Munro}}}]{strict1}
\bibinfo{author}{\bibfnamefont{G.~C.} \bibnamefont{{Knee}}},
  \bibinfo{author}{\bibfnamefont{K.}~\bibnamefont{{Kakuyanagi}}},
  \bibinfo{author}{\bibfnamefont{M.-C.} \bibnamefont{{Yeh}}},
  \bibinfo{author}{\bibfnamefont{Y.}~\bibnamefont{{Matsuzaki}}},
  \bibinfo{author}{\bibfnamefont{H.}~\bibnamefont{{Toida}}},
  \bibinfo{author}{\bibfnamefont{H.}~\bibnamefont{{Yamaguchi}}},
  \bibinfo{author}{\bibfnamefont{S.}~\bibnamefont{{Saito}}},
  \bibinfo{author}{\bibfnamefont{A.~J.} \bibnamefont{{Leggett}}},
  \bibnamefont{and} \bibinfo{author}{\bibfnamefont{W.~J.}
  \bibnamefont{{Munro}}}, \bibinfo{journal}{Nature Communications}
  \textbf{\bibinfo{volume}{7}}, \bibinfo{eid}{13253} (\bibinfo{year}{2016}),
  \eprint{1601.03728}.

\bibitem[{\citenamefont{{Liu} et~al.}(2019)\citenamefont{{Liu}, {Zhou}, {Han},
  {Li}, {Hu}, {Yang}, {Li}, {Liu}, {Li}, {Ma} et~al.}}]{strict2}
\bibinfo{author}{\bibfnamefont{X.}~\bibnamefont{{Liu}}},
  \bibinfo{author}{\bibfnamefont{Z.-Q.} \bibnamefont{{Zhou}}},
  \bibinfo{author}{\bibfnamefont{Y.-J.} \bibnamefont{{Han}}},
  \bibinfo{author}{\bibfnamefont{Z.-F.} \bibnamefont{{Li}}},
  \bibinfo{author}{\bibfnamefont{J.}~\bibnamefont{{Hu}}},
  \bibinfo{author}{\bibfnamefont{T.-S.} \bibnamefont{{Yang}}},
  \bibinfo{author}{\bibfnamefont{P.-Y.} \bibnamefont{{Li}}},
  \bibinfo{author}{\bibfnamefont{C.}~\bibnamefont{{Liu}}},
  \bibinfo{author}{\bibfnamefont{X.}~\bibnamefont{{Li}}},
  \bibinfo{author}{\bibfnamefont{Y.}~\bibnamefont{{Ma}}}, \bibnamefont{et~al.},
  \bibinfo{journal}{\pra} \textbf{\bibinfo{volume}{100}}, \bibinfo{eid}{032106}
  (\bibinfo{year}{2019}), \eprint{1910.10865}.

\bibitem[{\citenamefont{Halliwell}(2019)}]{Halliwell:2018dor}
\bibinfo{author}{\bibfnamefont{J.~J.} \bibnamefont{Halliwell}},
  \bibinfo{journal}{J. Phys. Conf. Ser.} \textbf{\bibinfo{volume}{1275}},
  \bibinfo{pages}{012008} (\bibinfo{year}{2019}), \eprint{1811.10408}.

\bibitem[{\citenamefont{Weisskopf and
  Wigner}(1930{\natexlab{a}})}]{Weisskopf:1930au}
\bibinfo{author}{\bibfnamefont{V.}~\bibnamefont{Weisskopf}} \bibnamefont{and}
  \bibinfo{author}{\bibfnamefont{E.~P.} \bibnamefont{Wigner}},
  \bibinfo{journal}{Z. Phys.} \textbf{\bibinfo{volume}{63}},
  \bibinfo{pages}{54} (\bibinfo{year}{1930}{\natexlab{a}}).

\bibitem[{\citenamefont{Weisskopf and
  Wigner}(1930{\natexlab{b}})}]{Weisskopf:1930ps}
\bibinfo{author}{\bibfnamefont{V.}~\bibnamefont{Weisskopf}} \bibnamefont{and}
  \bibinfo{author}{\bibfnamefont{E.}~\bibnamefont{Wigner}},
  \bibinfo{journal}{Z. Phys.} \textbf{\bibinfo{volume}{65}},
  \bibinfo{pages}{18} (\bibinfo{year}{1930}{\natexlab{b}}).

\bibitem[{\citenamefont{Fonda et~al.}(1978)\citenamefont{Fonda, Ghirardi, and
  Rimini}}]{Fonda:1978dk}
\bibinfo{author}{\bibfnamefont{L.}~\bibnamefont{Fonda}},
  \bibinfo{author}{\bibfnamefont{G.~C.} \bibnamefont{Ghirardi}},
  \bibnamefont{and} \bibinfo{author}{\bibfnamefont{A.}~\bibnamefont{Rimini}},
  \bibinfo{journal}{Rept. Prog. Phys.} \textbf{\bibinfo{volume}{41}},
  \bibinfo{pages}{587} (\bibinfo{year}{1978}).

\bibitem[{\citenamefont{Wilkinson
  et~al.}(1997{\natexlab{a}})\citenamefont{Wilkinson, Bharucha, Fischer,
  Madison, Morrow, Niu, Sundaram, and Raizen}}]{raizen1}
\bibinfo{author}{\bibfnamefont{S.~R.} \bibnamefont{Wilkinson}},
  \bibinfo{author}{\bibfnamefont{C.~F.} \bibnamefont{Bharucha}},
  \bibinfo{author}{\bibfnamefont{M.~C.} \bibnamefont{Fischer}},
  \bibinfo{author}{\bibfnamefont{K.~W.} \bibnamefont{Madison}},
  \bibinfo{author}{\bibfnamefont{P.~R.} \bibnamefont{Morrow}},
  \bibinfo{author}{\bibfnamefont{Q.}~\bibnamefont{Niu}},
  \bibinfo{author}{\bibfnamefont{B.}~\bibnamefont{Sundaram}}, \bibnamefont{and}
  \bibinfo{author}{\bibfnamefont{M.~G.} \bibnamefont{Raizen}},
  \bibinfo{journal}{Nature} \textbf{\bibinfo{volume}{387}},
  \bibinfo{pages}{575} (\bibinfo{year}{1997}{\natexlab{a}}).

\bibitem[{\citenamefont{{Fischer} et~al.}(2001)\citenamefont{{Fischer},
  {Guti{\'e}rrez-Medina}, and {Raizen}}}]{raizen2}
\bibinfo{author}{\bibfnamefont{M.~C.} \bibnamefont{{Fischer}}},
  \bibinfo{author}{\bibfnamefont{B.}~\bibnamefont{{Guti{\'e}rrez-Medina}}},
  \bibnamefont{and} \bibinfo{author}{\bibfnamefont{M.~G.}
  \bibnamefont{{Raizen}}}, \bibinfo{journal}{Phys. Rev. Lett.}
  \textbf{\bibinfo{volume}{87}}, \bibinfo{eid}{040402} (\bibinfo{year}{2001}),
  \eprint{quant-ph/0104035}.

\bibitem[{\citenamefont{Rothe et~al.}(2006)\citenamefont{Rothe, Hintschich, and
  Monkman}}]{rothe}
\bibinfo{author}{\bibfnamefont{C.}~\bibnamefont{Rothe}},
  \bibinfo{author}{\bibfnamefont{S.~I.} \bibnamefont{Hintschich}},
  \bibnamefont{and} \bibinfo{author}{\bibfnamefont{A.~P.}
  \bibnamefont{Monkman}}, \bibinfo{journal}{Phys. Rev. Lett.}
  \textbf{\bibinfo{volume}{96}}, \bibinfo{pages}{163601}
  (\bibinfo{year}{2006}).

\bibitem[{\citenamefont{Crespi et~al.}(2019)\citenamefont{Crespi, Pepe, Facchi,
  Sciarrino, Mataloni, Nakazato, Pascazio, and Osellame}}]{lastpasca}
\bibinfo{author}{\bibfnamefont{A.}~\bibnamefont{Crespi}},
  \bibinfo{author}{\bibfnamefont{F.~V.} \bibnamefont{Pepe}},
  \bibinfo{author}{\bibfnamefont{P.}~\bibnamefont{Facchi}},
  \bibinfo{author}{\bibfnamefont{F.}~\bibnamefont{Sciarrino}},
  \bibinfo{author}{\bibfnamefont{P.}~\bibnamefont{Mataloni}},
  \bibinfo{author}{\bibfnamefont{H.}~\bibnamefont{Nakazato}},
  \bibinfo{author}{\bibfnamefont{S.}~\bibnamefont{Pascazio}}, \bibnamefont{and}
  \bibinfo{author}{\bibfnamefont{R.}~\bibnamefont{Osellame}},
  \bibinfo{journal}{Phys. Rev. Lett.} \textbf{\bibinfo{volume}{122}},
  \bibinfo{pages}{130401} (\bibinfo{year}{2019}).

\bibitem[{\citenamefont{Degasperis et~al.}(1973)\citenamefont{Degasperis,
  Fonda, and Ghirardi}}]{Degasperis:1973wce}
\bibinfo{author}{\bibfnamefont{A.}~\bibnamefont{Degasperis}},
  \bibinfo{author}{\bibfnamefont{L.}~\bibnamefont{Fonda}}, \bibnamefont{and}
  \bibinfo{author}{\bibfnamefont{G.~C.} \bibnamefont{Ghirardi}},
  \bibinfo{journal}{Nuovo Cim. A} \textbf{\bibinfo{volume}{21}},
  \bibinfo{pages}{471} (\bibinfo{year}{1973}).

\bibitem[{\citenamefont{Misra and Sudarshan}(1977)}]{Misra:1976by}
\bibinfo{author}{\bibfnamefont{B.}~\bibnamefont{Misra}} \bibnamefont{and}
  \bibinfo{author}{\bibfnamefont{E.~C.~G.} \bibnamefont{Sudarshan}},
  \bibinfo{journal}{J. Math. Phys.} \textbf{\bibinfo{volume}{18}},
  \bibinfo{pages}{756} (\bibinfo{year}{1977}).

\bibitem[{\citenamefont{Itano et~al.}(1990)\citenamefont{Itano, Heinzen,
  Bollinger, and Wineland}}]{Itano:1990zz}
\bibinfo{author}{\bibfnamefont{W.~M.} \bibnamefont{Itano}},
  \bibinfo{author}{\bibfnamefont{D.~J.} \bibnamefont{Heinzen}},
  \bibinfo{author}{\bibfnamefont{J.~J.} \bibnamefont{Bollinger}},
  \bibnamefont{and} \bibinfo{author}{\bibfnamefont{D.~J.}
  \bibnamefont{Wineland}}, \bibinfo{journal}{Phys. Rev.}
  \textbf{\bibinfo{volume}{A41}}, \bibinfo{pages}{2295} (\bibinfo{year}{1990}).

\bibitem[{\citenamefont{Balzer et~al.}(2000)\citenamefont{Balzer, Huesmann,
  Neuhauser, and Toschek}}]{balzer}
\bibinfo{author}{\bibfnamefont{C.}~\bibnamefont{Balzer}},
  \bibinfo{author}{\bibfnamefont{R.}~\bibnamefont{Huesmann}},
  \bibinfo{author}{\bibfnamefont{W.}~\bibnamefont{Neuhauser}},
  \bibnamefont{and} \bibinfo{author}{\bibfnamefont{P.~E.}
  \bibnamefont{Toschek}}, \bibinfo{journal}{Opt. Comm. 180}
  \textbf{\bibinfo{volume}{180}}, \bibinfo{pages}{115} (\bibinfo{year}{2000}).

\bibitem[{\citenamefont{{Streed} et~al.}(2006)\citenamefont{{Streed}, {Mun},
  {Boyd}, {Campbell}, {Medley}, {Ketterle}, and {Pritchard}}}]{streed}
\bibinfo{author}{\bibfnamefont{E.~W.} \bibnamefont{{Streed}}},
  \bibinfo{author}{\bibfnamefont{J.}~\bibnamefont{{Mun}}},
  \bibinfo{author}{\bibfnamefont{M.}~\bibnamefont{{Boyd}}},
  \bibinfo{author}{\bibfnamefont{G.~K.} \bibnamefont{{Campbell}}},
  \bibinfo{author}{\bibfnamefont{P.}~\bibnamefont{{Medley}}},
  \bibinfo{author}{\bibfnamefont{W.}~\bibnamefont{{Ketterle}}},
  \bibnamefont{and} \bibinfo{author}{\bibfnamefont{D.~E.}
  \bibnamefont{{Pritchard}}}, \bibinfo{journal}{\prl}
  \textbf{\bibinfo{volume}{97}}, \bibinfo{eid}{260402} (\bibinfo{year}{2006}),
  \eprint{cond-mat/0606430}.

\bibitem[{\citenamefont{Kofman and Kurizki}(2000)}]{KK1}
\bibinfo{author}{\bibfnamefont{A.}~\bibnamefont{Kofman}} \bibnamefont{and}
  \bibinfo{author}{\bibfnamefont{G.}~\bibnamefont{Kurizki}},
  \bibinfo{journal}{Nature} \textbf{\bibinfo{volume}{405}},
  \bibinfo{pages}{546} (\bibinfo{year}{2000}).

\bibitem[{\citenamefont{Kofman and Kurizki}(2001)}]{KK2}
\bibinfo{author}{\bibfnamefont{A.}~\bibnamefont{Kofman}} \bibnamefont{and}
  \bibinfo{author}{\bibfnamefont{G.}~\bibnamefont{Kurizki}},
  \bibinfo{journal}{Zeitschrift für Naturforschung A}
  \textbf{\bibinfo{volume}{56}} (\bibinfo{year}{2001}).

\bibitem[{\citenamefont{{Facchi}
  et~al.}(2001{\natexlab{a}})\citenamefont{{Facchi}, {Nakazato}, and
  {Pascazio}}}]{prlpasca}
\bibinfo{author}{\bibfnamefont{P.}~\bibnamefont{{Facchi}}},
  \bibinfo{author}{\bibfnamefont{H.}~\bibnamefont{{Nakazato}}},
  \bibnamefont{and}
  \bibinfo{author}{\bibfnamefont{S.}~\bibnamefont{{Pascazio}}},
  \bibinfo{journal}{Phys. Rev. Lett.} \textbf{\bibinfo{volume}{86}},
  \bibinfo{pages}{2699} (\bibinfo{year}{2001}{\natexlab{a}}),
  \eprint{quant-ph/0006094}.

\bibitem[{\citenamefont{{Facchi}
  et~al.}(2001{\natexlab{b}})\citenamefont{{Facchi}, {Nakazato}, and
  {Pascazio}}}]{FPprl}
\bibinfo{author}{\bibfnamefont{P.}~\bibnamefont{{Facchi}}},
  \bibinfo{author}{\bibfnamefont{H.}~\bibnamefont{{Nakazato}}},
  \bibnamefont{and}
  \bibinfo{author}{\bibfnamefont{S.}~\bibnamefont{{Pascazio}}},
  \bibinfo{journal}{Phys. Rev. Lett.} \textbf{\bibinfo{volume}{86}},
  \bibinfo{pages}{2699} (\bibinfo{year}{2001}{\natexlab{b}}),
  \eprint{quant-ph/0006094}.

\bibitem[{\citenamefont{Giacosa and Pagliara}(2020)}]{Giacosa:2019nbz}
\bibinfo{author}{\bibfnamefont{F.}~\bibnamefont{Giacosa}} \bibnamefont{and}
  \bibinfo{author}{\bibfnamefont{G.}~\bibnamefont{Pagliara}},
  \bibinfo{journal}{Phys. Rev. D} \textbf{\bibinfo{volume}{101}},
  \bibinfo{pages}{056003} (\bibinfo{year}{2020}), \eprint{1906.10024}.

\bibitem[{\citenamefont{Koshino and Shimizu}(2005)}]{Koshino:2004rw}
\bibinfo{author}{\bibfnamefont{K.}~\bibnamefont{Koshino}} \bibnamefont{and}
  \bibinfo{author}{\bibfnamefont{A.}~\bibnamefont{Shimizu}},
  \bibinfo{journal}{Phys. Rept.} \textbf{\bibinfo{volume}{412}},
  \bibinfo{pages}{191} (\bibinfo{year}{2005}), \eprint{quant-ph/0411145}.

\bibitem[{\citenamefont{Wilkinson
  et~al.}(1997{\natexlab{b}})\citenamefont{Wilkinson, Bharucha, Fischer,
  Madison, Morrow, Niu, Sundaram, and Raizen}}]{Reizen1}
\bibinfo{author}{\bibfnamefont{S.~R.} \bibnamefont{Wilkinson}},
  \bibinfo{author}{\bibfnamefont{C.~F.} \bibnamefont{Bharucha}},
  \bibinfo{author}{\bibfnamefont{M.~C.} \bibnamefont{Fischer}},
  \bibinfo{author}{\bibfnamefont{K.~W.} \bibnamefont{Madison}},
  \bibinfo{author}{\bibfnamefont{P.~R.} \bibnamefont{Morrow}},
  \bibinfo{author}{\bibfnamefont{Q.}~\bibnamefont{Niu}},
  \bibinfo{author}{\bibfnamefont{B.}~\bibnamefont{Sundaram}}, \bibnamefont{and}
  \bibinfo{author}{\bibfnamefont{M.~G.} \bibnamefont{Raizen}},
  \bibinfo{journal}{Nature} \textbf{\bibinfo{volume}{387}},
  \bibinfo{pages}{575} (\bibinfo{year}{1997}{\natexlab{b}}).

\bibitem[{\citenamefont{Bassi et~al.}(2013)\citenamefont{Bassi, Lochan, Satin,
  Singh, and Ulbricht}}]{Bassi:2012bg}
\bibinfo{author}{\bibfnamefont{A.}~\bibnamefont{Bassi}},
  \bibinfo{author}{\bibfnamefont{K.}~\bibnamefont{Lochan}},
  \bibinfo{author}{\bibfnamefont{S.}~\bibnamefont{Satin}},
  \bibinfo{author}{\bibfnamefont{T.~P.} \bibnamefont{Singh}}, \bibnamefont{and}
  \bibinfo{author}{\bibfnamefont{H.}~\bibnamefont{Ulbricht}},
  \bibinfo{journal}{Rev. Mod. Phys.} \textbf{\bibinfo{volume}{85}},
  \bibinfo{pages}{471} (\bibinfo{year}{2013}), \eprint{1204.4325}.

\bibitem[{\citenamefont{{Facchi} and {Pascazio}}(2001)}]{factorzeta}
\bibinfo{author}{\bibfnamefont{P.}~\bibnamefont{{Facchi}}} \bibnamefont{and}
  \bibinfo{author}{\bibfnamefont{S.}~\bibnamefont{{Pascazio}}},
  \bibinfo{journal}{arXiv e-prints} \bibinfo{eid}{quant-ph/0101044}
  (\bibinfo{year}{2001}), \eprint{quant-ph/0101044}.

\bibitem[{\citenamefont{Khalfin}(1957)}]{khalfin}
\bibinfo{author}{\bibfnamefont{L.~A.} \bibnamefont{Khalfin}},
  \bibinfo{journal}{Zh. Eksp. Q Teor. Fiz} \textbf{\bibinfo{volume}{33}}
  (\bibinfo{year}{1957}).

\bibitem[{\citenamefont{{Facchi} and {Pascazio}}(1998)}]{FPatoms}
\bibinfo{author}{\bibfnamefont{P.}~\bibnamefont{{Facchi}}} \bibnamefont{and}
  \bibinfo{author}{\bibfnamefont{S.}~\bibnamefont{{Pascazio}}},
  \bibinfo{journal}{Physics Letters A} \textbf{\bibinfo{volume}{241}},
  \bibinfo{pages}{139} (\bibinfo{year}{1998}), \eprint{quant-ph/9905017}.

\bibitem[{\citenamefont{Giacosa et~al.}(2020)\citenamefont{Giacosa, Ko\'scik,
  and Sowi\'nski}}]{Giacosa:2019jxz}
\bibinfo{author}{\bibfnamefont{F.}~\bibnamefont{Giacosa}},
  \bibinfo{author}{\bibfnamefont{P.}~\bibnamefont{Ko\'scik}}, \bibnamefont{and}
  \bibinfo{author}{\bibfnamefont{T.}~\bibnamefont{Sowi\'nski}},
  \bibinfo{journal}{Phys. Rev. A} \textbf{\bibinfo{volume}{102}},
  \bibinfo{pages}{022204} (\bibinfo{year}{2020}), \eprint{1912.06394}.

\bibitem[{\citenamefont{Giacosa}(2012)}]{Giacosa:2011xa}
\bibinfo{author}{\bibfnamefont{F.}~\bibnamefont{Giacosa}},
  \bibinfo{journal}{Found. Phys.} \textbf{\bibinfo{volume}{42}},
  \bibinfo{pages}{1262} (\bibinfo{year}{2012}), \eprint{1110.5923}.

\bibitem[{\citenamefont{Giacosa}(2021)}]{Giacosa:2021hgl}
\bibinfo{author}{\bibfnamefont{F.}~\bibnamefont{Giacosa}}
  (\bibinfo{year}{2021}), \eprint{2108.07838}.

\end{thebibliography}
\end{document}